\documentclass[12pt]{article}
\usepackage{graphicx}

\usepackage{amssymb}
\usepackage{amsmath}
\usepackage{amsfonts}

\def\Xv{{\bf X}}

\def\bv{{\bf b}}

\def\kv{{\bf k}}

\def\pv{{\bf p}}
\def\qv{{\bf q}}
\def\rv{{\bf r}}

\def\vv{{\bf v}}

\def\Av{{\bf A}}
\def\Bv{{\bf B}}

\def\Ev{{\bf E}}
\def\Fv{{\bf F}}

\def\Rv{{\bf R}}

\def\nv{{\bf n}}

\def\be{\begin{equation}}
\def\ee{\end{equation}}
\def\ni{\noindent}

\def\lambdabar{\lambda\raise0.4ex\hbox{\kern-0.5em\hbox{--}}\ }
\def\lambdaC{\lambda\raise0.5ex\hbox{\kern-0.5em\hbox{--}}_{\rm C}}
\def\lambdabarc{\lambda\raise0.5ex\hbox{\kern-0.5em\hbox{--}}_{\rm c}}
\def\lesssim{\,{\lower0.5ex\hbox{$\stackrel{<}{\sim}$}}\,}
\def\gtrsim{\,{\lower0.5ex\hbox{$\stackrel{>}{\sim}$}}\,}

\def\me{m_{\rm e}}
\def\td{t_{\rm d}}
\def\CVR{${\rm \check CVR}$}

\def\energ{\epsilon}
\def\t{_{\perp}}
\def\l{_{\parallel}}
\def\T{_{\rm T}}
\def\L{_{\rm L}}
\def\pl{_{\rm P}}
\def\epol{{\bf{\hat e}}}
\def\ccdot{\!\cdot\!}

\begin{document}


\title{Classical and Quantum Phenomenology \\ in Radiation by Relativistic Electrons \\ in Matter or in External Fields}

\maketitle     


\begin{center}

\medskip
\author{Xavier Artru}   

Universit\'e de Lyon, Institut de Physique Nucl\'eaire de Lyon,  \\Universit\'e Lyon~1 and CNRS-IN2P3, France



\bigskip

In memory of Vladimir Nikolaevich Baier \\ and Vladimir Moisevich Strakhovenko

\end{center}

\begin{abstract}

Phenomenological aspects of radiation by relativistic electrons in external field, in matter or the vicinity of matter are reviewed, among which: infrared divergence,  coherence length effects, 
shadowing, enhancement in aligned crystals, quantum recoil and spin effects, electron side-slipping, photon impact parameter and the presence of tunnelling in the radiation process. 
\end{abstract}



\section{Introduction}

In Classical Electrodynamics (CED), a moving electron can emit radiation by two mechanisms: 

\paragraph{(A) velocity change :}the photon is directly emitted from the electron world line. This is the case of \emph{Synchrotron Radiation} (SR), \emph{Undulator Radiation} (UR), \emph{Compton Back-Scattering} (CBS), \emph{Bremsstrahlung} (BR), \emph{Coherent Bremsstrahlung} (CBR) and \emph{Channeling Radiation}Ê(CR).
  
\paragraph{(B) transient medium polarization :}
 If the motion is \emph{balistic}  (i.e., rectilinear and uniform),  but inside or near a medium, the photon is emitted by the polarization currents induced by the travelling Coulomb field of the electron. This is the case of 
\emph{Cherenkov-Vavilov Radiation} (\CVR), \emph{Transition Radiation} (TR) in optical (OTR) or X-ray (XTR) domains, 
\emph{Diffraction Radiation} (DR), \emph{Smith-Purcell Radiation} (SPR), \emph{Parametric X-Rays} (PXR) and \emph{Polarization Bremsstrahlung} (PBR) on individual atoms. 

 (A) and (B) can coexist, e.g., in Diffracted Channeling Radiation. 

 \medskip\ni 
 A  vast theoretical and experimental work is currently done on the above-listed radiations, specially after the prediction of intense Channeling Radiation by relativistic electrons by Muradin Kumakhov \cite{KUMA}.
  In this paper we review some of their ``universal" properties in a phenomenological approach. We gather these properties under three headings:

\ni - {\bf Classical radiation} (Section 2), for photon energy $\omega$ much smaller than the electron energy $\energ$~;
 
\ni - {\bf Hard photon emission} (Section 3), for $\omega\sim\energ$~; 

\ni - {\bf Impact parameter description} (Section 4). 

\ni We will assume that the electron is ultrarelativistic ($\energ/\me\equiv\gamma\gg1$) and classical, at least between photon emissions. 
We will not speak about its dynamics in the "radiator" (UR device, amorphous matter, crystal, TR or PXR targets, etc.) which, in crystals, involves many phenomena~: channeling, dechanneling, volume capture, volume reflection, etc. We suppose that the trajectory of the electron has been calculated beforehand. For the general theory of radiation the reader may consult \cite{PANOVSKY,JACKSON,GINZBURG} and for BR, CR and CBR, \cite{BELO-KOMA,SHULGA,BAIER,SHULGA'}. Many considerations presented below can be found in \cite{RAD, XA-Berlin}.

We will work with natural unit systems where $\hbar=c=1$; $\alpha={e^2}/({4\pi})\simeq {1/137}$; the letter "$\gamma$" can also designate a photon, of momentum $\kv=\omega\,\nv$; $n=|\nv|$ is the refraction index. 
The components $X\l$  and $\Xv\t$ of a vector $\Xv$ are parallel and perpendicular to $\kv$, whereas $X\L$  and $\Xv\T$ are relative to the electron velocity $\vv(t)$, or to $\langle\vv\rangle$ for CR and CBR.   

\section{Classical radiation}
We consider an electron of classical trajectory $\rv=\rv(t)$ and suppose that the emission of one photon does not modify the trajectory noticeably and does not involve the electron spin. This excludes channeling at low energy ($\energ\lesssim100$ MeV), where the number of transverse energy states is low (quantum channeling regime) and at very high energy, where hard photon emission ($\omega\sim\energ$) takes place. For a given polarization $\hat e$, the photon spectrum writes
\be\label{cov1} 
dN(\hat e) = \frac{\alpha}{4\pi^2} \, \frac{d^3\kv}{\omega} \, \left|{\cal A}\right|^2 \,.
\ee
For mechanism (A) in vacuum, ${\cal A}$ is given in covariant way by
%
\be\label{cov2}     
{\cal A} = - \hat e^* \cdot A(k)\,, \quad 
A = \int_{\cal T} dX \, \exp(i\phi)\,, \quad \phi = k\cdot X \,.
\ee
$X=(t,\rv)$, $k=(\omega,\kv)$, $A$ and $\hat e$ 
are 4-vectors. We take the metric where $k\ccdot X= \omega t - \kv\ccdot\rv$, $\hat e^*\ccdot\hat e=-1$. The integral in (\ref{cov2}) is along the whole electron trajectory ${\cal T}$. In a non-covariant formulation we take the gauge $\hat e^0=\epol\perp\kv$, $\epol^*\ccdot\epol$, and replace $-\hat e^* \ccdot A$ by $\epol^* \ccdot \Av\t$ with
%
\be\label{noncov} 
 \Av =  \int_{\cal T} d\td \exp(i\phi) 
 \, \frac{d\rv}{d\td}
=   \frac{i}{\omega} \int_{\cal T} d\td \exp(i\phi) \,  \frac{d^2\rv}{d\td^2} \,, \quad \phi=\omega \td \,.
\ee
$\td=t-\nv\ccdot\rv$ is the \emph{detection time} up to a constant; 
$d\rv\t / d\td$ and $d^2\rv\t / d\td^2$ are the \emph{apparent} perpendicular velocity and acceleration. There is no bound on $|d\rv\t / d\td|$. The second expression of $\Av$ emphasizes the role of the acceleration. 

 From the QED point of view, the emitted photons form a \emph{coherent state}. The number of photons in any given $\kv$ domain has a Poisson distribution.

\medskip
For mechanism (B), or a coexistence of (A) and (B), ${\cal A}$  can be calculated using the reciprocity theorem, related to time reversal. It gives \cite{RAD}
\be\label{recip} 
 {\cal A} 
=  \int_{\rm revers({\cal T})}  d\rv\cdot \Ev^{\rm (in)}_{-\kv,\epol^*}(t,\rv)
\,,  
\ee
where revers(${\cal T}$) is the {time-reversed trajectory} of the electron. 
$\Ev^{\rm (in)}_{-\kv,\epol^*}$
is the complex electric field of the \emph{ingoing} solution of the homogenous Maxwell equations in matter, for a wave coming from the detector with momentum $-\kv$ and polarization $\epol^*$. It is normalized to 
$\epol^* \exp(-i\omega t - i\kv\ccdot\rv)$ in vacuum
in the detector direction. Equation (\ref{recip}) is well suited to TR, SPR and PXR. With a complex refraction index $n$, it takes the absorption in the radiator into account. 

In the following we will make the ultrarelativistic and small-angle approximations $\gamma\gg1$,  $|\vv\T(t)|$ and $|\nv\T| \ll1$. In the X-ray domain, $1-|\nv|\simeq\omega\pl^2/(2\omega^2)\ll1$,  $\omega\pl$ being the plasma frequency, and   
\be \label{dtau}
d\td/dt 
\simeq (\gamma^{-2}+\theta^2+\omega\pl^2/\omega^2)/2 \ll 1\,,
\ee
%
with $\vec\theta\equiv-\vv\t(t) \simeq \nv-\vv(t) 
$. Equation (\ref{noncov}), supplemented by (\ref{dtau}), also applies to XTR, neglecting the refractions of the X-ray. 

\paragraph{Sudden velocity change.} 
As a prototype of mechanism (A), we consider an electron of trajectory $\rv=\vv_{\rm I} t$ for $t<0$ and $\rv=\vv_{\rm F} t$ for $t>0$ in vacuum. Then, 
%
\be\label{DIR2} 
\Av\t =\Av\t(\vv_{\rm I},\kv)-\Av\t(\vv_{\rm F},\kv) \,,
\ee
with
\be\label{lobe1} 
\Av\t(\vv,\kv) =
(i\omega)^{-1} \, \vv\t \, / (1-\nv\cdot\vv) 
\simeq (2i/\omega) \, {\vec\theta}\,/({\gamma^{-2}+\theta^2}) \,. 
\ee
%
For $\vv_{\rm F}=0$ we have a \emph{suddenly stopped} electron, for $\vv_{\rm I}=0$  a \emph{suddenly starting} electron. They emit the same photon spectrum,
\be\label{lobe2} 
dN(\vec\varepsilon)  = \frac{\alpha}{\pi^2} \, \frac{d\omega}{\omega} \, d\Omega \left|\frac{\vec\theta\cdot\vec\varepsilon}{\gamma^{-2}+\theta^2}\right|^2 \,,
\ee
but with opposite amplitudes. It can be understood with the superposition principle:  
to a suddenly stopping $e^+$, we can add - in thought - an $e^-$ comoving with the $e^+$ at $t<0$, but not stopping. This addition does not change the radiation, and the new system is equivalent to a suddenly starting $e^-$.

 The spectrum (\ref{lobe2}) has the following "universal" properties~:
\begin{itemize}
\item an annular-lobe angular distribution peaked at $\theta=1/\gamma$,

\item a radial polarization (\emph{i.e.,} in the $(\vv,\nv)$ plane),

\item an infrared divergence $dN/(d\omega d\Omega) \sim \omega^{-1}$, related to the semi-infinite balistic motion, 

\item an ultraviolet divergence, related to the infinite acceleration at $t=0$. It does not happen really, because $\omega< \energ$.

\end{itemize}
The case $|\vv_{\rm I}|=|\vv_{\rm F}|$ corresponds to Bremsstrahlung on one atom, in the Born approximation.
At large scattering angle ($|\vv_{\rm I}-\vv_{\rm F}| \gg \gamma^{-1}$)
we have two well-separated annular lobes. For the so-called \emph{dipolar} regime $|\vv_{\rm I}-\vv_{\rm F}| \ll \gamma^{-1}$, the interference between the lobes is essential. 

Whether the velocity change is sudden or not, Eqs.(\ref{DIR2},\ref{lobe1}) apply in the infrared part of the spectrum.  It suffices that the motion is balistic for $t\to\pm\infty$. The annular lobe (\ref{lobe2}) appears in many other experimental situations: in backward OTR from a metallic foil, in forward OTR from any screen, in the \emph{edge radiation} from a bending magnet, etc., although the finite size of the machine provides an infrared cutoff. 

In simulations, one truncates the integral in (\ref{noncov}). If we do it in the second expression, we introduce a spurious infrared divergence.  
It is like completing the trajectories by semi-infinite balistic motions. A phenomenological infrared cutoff has to be applied. In \cite{XA-code} such a cutoff was chosen with the help of a filtered sum rule \cite{XA-Elbruz}. Truncating in the first expression of  (\ref{noncov}) produces a spurious ultraviolet divergence, as if the electron started and stopped suddenly.   

\paragraph{Longitudinal coherence (LC) \cite{TER-MIK}.}

We define the {\it coherence lenght} as the length of a segment of trajectory over which $\phi$ increases by $\Delta\phi=1$ radian. In the X-ray domain and for a straight trajectory in uniform medium,
\be \label{Lcoh}
L_{\rm coh}(\vv,\kv)  = dt/d\phi 
\simeq (2/\omega) \, (\gamma^{-2}+\theta^2+\omega\pl^2/\omega^2)^{-1} \,.
\ee
It can reach macroscopic values, even though the photon wavelength is microscopic. 
For a \emph{curved} trajectory, the $\Delta\phi=1$ arcs have various lengths. We choose the longest one (where $\theta$ is minimum).  It gives
%
\be \label{Lcohcirc}
L_{\rm coh}  \simeq \min \{(2/\omega) \, (\gamma^{-2}+\psi^2)^{-1} \,, (24R^2/\omega)^{1/3} \} \,,
\ee
$\psi$ being angle between the photon and the curvature plane.

If the electron undergoes two successive scattering $\vv_{\rm I}\to\vv$ at $X=(t,\rv)$ and $\vv\to\vv_{\rm F}$ at $X'=(t',\rv')$, the $XX'$ path contributes to the first form of (\ref{noncov}) by 
\be\label{DIR13} 
\Av^{[XX']} = \exp[i\phi(t)] \, (e^{i\Delta\phi}-1) 
\, \Av(\vv,\kv) \,, 
\ee
with $\Delta\phi  = L /L_{\rm coh}(\vv,\kv)$ and $L=|\rv'-\rv|$.
When $\Delta\phi <1$, $\Av^{[XX']}$ is suppressed by the factor $(e^{i\Delta\phi}-1)$, which kills the infrared divergence of  $\Av(\vv,\kv)$. This is one part of the explanation of the Landau-Pomeranchuk-Migdal (LPM) effect.  
Similarly, when the electron crosses two radiators 
(\emph{e.g.,} a bending magnet and a OTR mirror)
distant by $L\lesssim L_{\rm coh}$, the interference between the resulting radiations must be taken into account.  
LC is destructive in an XTR radiator when the foil spacing is less than $L_{\rm coh}^{\rm (vac.)}$, or when the foil thickness is less than $L_{\rm coh}^{\rm (med.)}$, as expected 
from a collective response of the atoms. 

In  a radiator of spatial period $\Lambda$, LC gathers the radiation spectrum in narrow ``resonance" peaks where $L_{\rm coh}=\Lambda/(2\pi\nu)$, $\nu$ being the harmonic number.
This \emph{interference of order} $\nu$  is at work in UR, in \emph{resonant} XTR and in CBR on atomic planes or coplanar atomic strings (``String-Of-String" radiation).  
From the quantum point of view, $L_{\rm coh}=-1/q\l$, where $\qv$ 
is the momentum brought by the radiator.  
For instance, CBR can be considered as  the virtual Compton process
\be \label{CBR-WW} 
 e^{-}(\energ,\pv) + \gamma_{\rm V}(0,\qv)
 \to e^{-}(\energ',\pv') + \gamma(\omega,\kv) \,,
\ee
where $ \gamma_{\rm V}(0,\qv)$ is a virtual photon of the crystal field and $\qv$ belongs to the reciprocal lattice.  
In Undulator Radiation, harmonics appear in the non-dipole regime 
$K\equiv\gamma^{2}\langle\vv\T^2\rangle\gtrsim1$, i.e., $\langle\pv\T^2\rangle\gtrsim m^2$. In this case we must replace $\theta^2$ by 
$\langle\vv\t^2\rangle=\theta^2+\langle\vv\T^2\rangle$ in (\ref{Lcoh}) and 
$\nu$ is the number of virtual photons, each of momentum $2\pi/\Lambda$, involved in the reaction. The external field then works beyond the Born approximation. 

Independently of its indented spectrum, CBR yields more energy than BR. This is because, when crossing one plane or string of atoms, the electron crosses  
 several atoms at nearly equal impact parameters $\bv_i$, therefore the associated Bremsstrahlung amplitudes add constructively, provided this occurs within a path $L\lesssim L_{\rm coh}$. One can then speak of a "zero-order" interference. 

 \paragraph{Typical frequency.}

Photons which mostly conribute to the radiated energy are such that $L_{\rm coh}\sim \Delta L$, where $\Delta L$ is the characteristic length of the variations of  the apparent transverse velocity $d\rv\t/d\td$. 
For an almost straight trajectory, it gives 
$\omega\sim \gamma^2 / (\Delta L)$. 
 For a circular trajectory, $\Delta L\sim R/\gamma$ and $\omega\sim \omega_c=\gamma^3/R$.

 \paragraph{Equivalent Photons, or Weizs\"acker-Williams (WW) Method.}

The vector $\Av\t(\vv,\kv)$ of Eq.(\ref{lobe1}) is, apart from a factor $(2/e)\exp({i\omega t})$, nearly equal to the Fourier transform 
\be\label{WW} 
\tilde\Ev\T(\vv,\kv)= ie \, \kv\T/( \kv\T^2+\gamma^{-2}k\L^2) \, \exp(-{i\omega t})
\ee
of the transverse component of the electron Coulomb field at $t=0$. According to the WW approach \cite{WW}, the latter evolves into the free radiation field (\ref{lobe1}) if the electron is suddenly stopped or scattered at large angle. The main difference between (\ref{lobe1}) and (\ref{WW}) concerns the dispersion relations, $\omega=|\kv|$ for the \emph{real} photons of (\ref{lobe1}) and $\omega=\vv\ccdot\kv$ for the \emph{virtual} photons of (\ref{WW}). This method explains in a simple way the 
spectrum and polarization of backward OTR, considered as a reflection of the equivalent photons. 

The WW method can also be applied to the virtual photons \emph{of the target}, if their virtuality $-q^2=\qv^2$ is not too high. CBR can thus be treated as Compton scattering (see Eq.\ref{CBR-WW}). The crossed process, Coherent Pair Creation on crystal planes, can be viewed as 
$ \gamma + \gamma_{\rm V}  \to e^{+} e^{-}$. 

In impact parameter space, the Fourier transform of (\ref{WW}) gives \cite{JACKSON}
\be\label{WWb} 
\Ev\T(k\L,\bv) \simeq - e \sqrt{2/\pi}Ê\, (\omega/\gamma) \,  K_1(\omega b/\gamma) \,.
\ee
It has a $1/b$ singularity at $b=0$ due to the slow decrease at $|\kv\T|\to\infty$. 
This singularity is not observed when focusing an optical system on an OTR target \cite{XA-OTR}, because of the limit $|\kv\T|<\omega\sin\theta_{\rm max}$, where $\theta_{\rm max}$ is the optical aperture. Instead one observes a \emph{dip} at $b=0$.

%
%

\paragraph{The semi-bare electron.}

A hard-scattered electron looses its initial photon cloud and is left quasi-bare \cite{FEINBERG,SHULGA-naked,BOLOTOVSKY}. The Coulomb field is gradually restored, starting with the short-distance, or high-momentum, components. For a given $\kv$, this process takes place in a \emph{formation zone} of length $L\sim L_{\rm coh}(\vv_{\rm F},\kv)$, while real photons are created, forming the annular lobe around $\vv_{\rm F}$. The same process happens when the electron leaves an opaque material, or in beta decay. If one puts a backward OTR or DR foil inside the formation zone, it will give a very low signal. This is another way to understand the destructive LC effect. 

\begin{figure} 
\begin{tabular}{cc}
\includegraphics*[angle=90, width=0.55\textwidth, bb= 70 135 460 715]{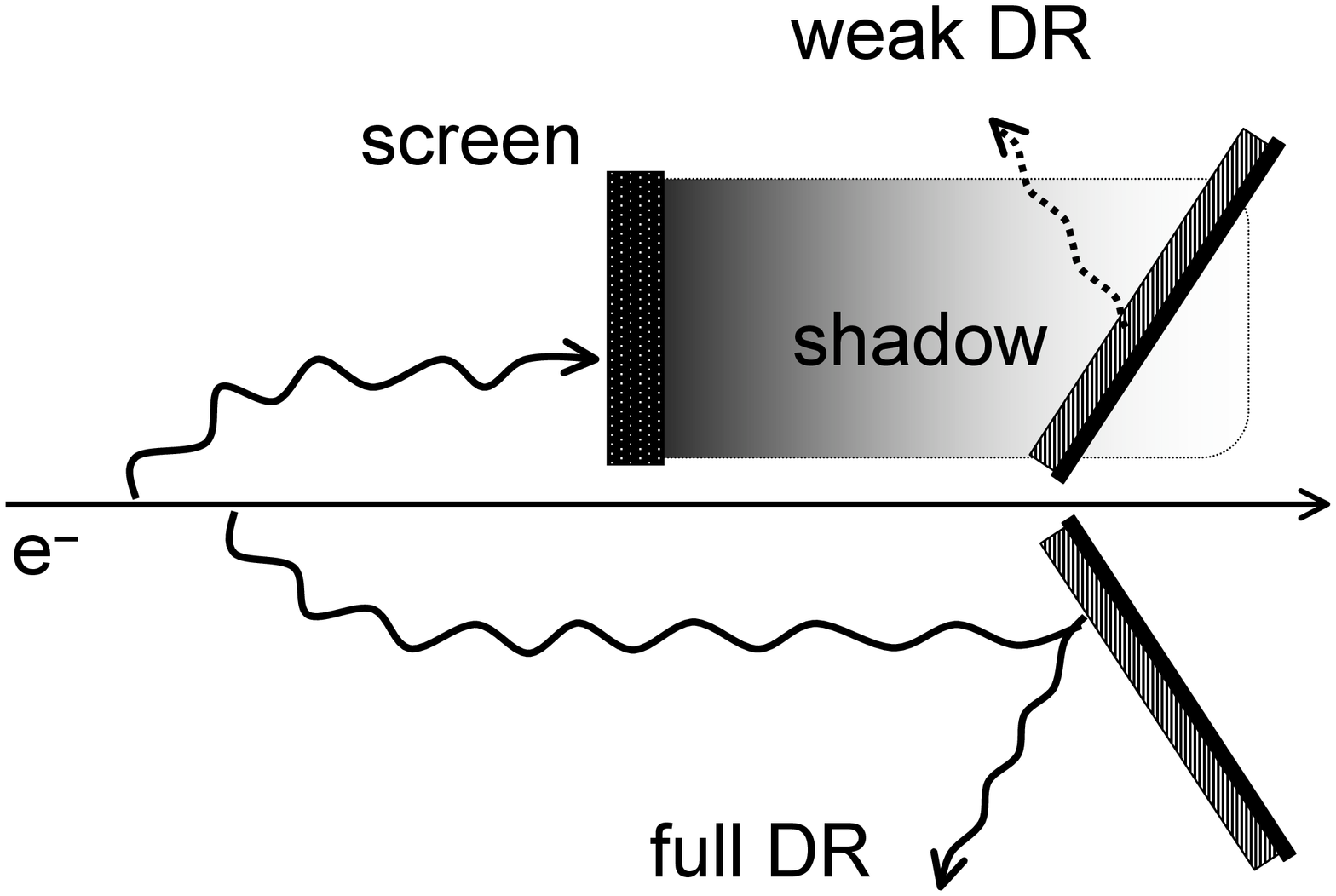}
&
\includegraphics*[angle=90, width=0.40\textwidth, bb= 100 200 485 600]{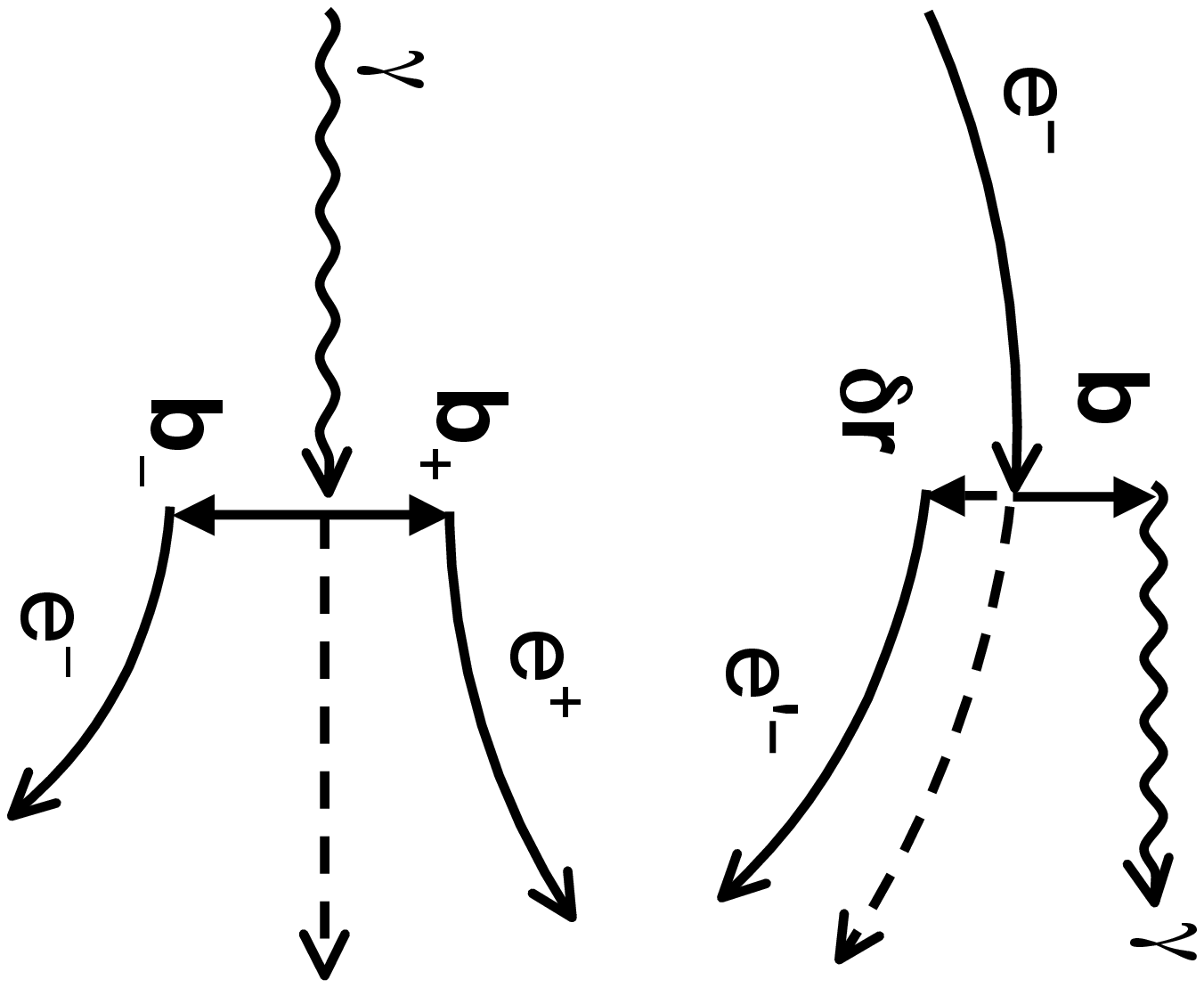}
%
\end{tabular}
\caption{\footnotesize Left : scheme of a shadowing experiment.
Right : side-slipping of the electron (upper diagram) and impact parameter of the photon (lower diagram).}
\end{figure}

\paragraph{Shadowing.} A screen whose contour is close to the electron trajectory intercepts part of the equivalent photon cloud, making a shadow in the formation zone  \cite{BOLOTOVSKY,XA-shadow}. This has been verified in an experiment at Tomsk  \cite{NAUM-XA} schematized in Fig.1, left. Shadowing limits in an essential way the Smith-Purcell radiation, since each ridge of the grating makes a shadow on the next one. A bound $dW/dL\le \alpha \, /(2\pi b^2)$ for the SPR energy per electron and per unit length has been proposed \cite{XA-bound}. 

\paragraph{Crystal-assisted radiation.}

It is weel known that an electron penetrating a crystal at small angle with major atomic planes or strings radiates more energy than at random orientation. In short, ``CBR and CR $>$ BR". It is interesting to have a qualitative explanation of that, without calculating the full CR or CBR spectrum. 

A heuristic explanation is that an electron crossing $N$ approximately aligned atoms in a relatively small path length $L$ emits Bremsstrahlung as on a "super-atom" of atomic number $NZ$, therefore $N^2$ times more than on one real atom. This is the  "zero-order" interference mentioned above. 
However such atoms are $N$ times less numerous than the real ones, therefore the net gain is only $N$. This number can be estimated by geometrical arguments, assuming  quasi-straight trajectories. 
By crossing an atomic string at small angle $\psi$,  $N\sim a/(d\psi)$, $a$ being the Thomas-Fermi radius and $d$ the lattice constant. 
By crossing an atomic plane,  $N\sim a^2/(d^2\psi)$. 
 In the channeling regime ($\psi\lesssim\psi_{\rm c}$) this calculation is not valid, but $N$ is still large. Thus we expect CBR to be $N$ times more intense than ordinary BR, at least for the frequencies such that $L= a/\psi \lesssim L_{\rm coh}$. If not, $N$ has to be reduced by the factor $L_{\rm coh}/L$.

On the other hand,  a classical sum rule predicts, at first sight, that the radiated energy is \emph{independent} of $\psi$. From (\ref{noncov}) one can derive various sum rules \cite{XA-Elbruz,RAD} for the photon spectra at fixed $\nv$, one of which is
\be \label{reglesom} 
\int d\omega \frac{\omega dN}{d\omega \, d\Omega} =  \frac{\alpha}{4\pi} \int d\td
\left(\frac{d^2\rv\t}{d\td^2}\right)^2 ;
\ee
integrating over $\nv$ gives the \emph{Li\'enard formula} for the radiated energy
\be \label{Larmor}
W=\frac{2\alpha}{3m^2} \int dt (F\L^2 + \gamma^2 \Fv\T^2) ,
\ee
$\Fv=\Fv[\rv(t)]$ being the force acting on the electron. Assuming that it comes from 
the microscopic  electric field $\Ev_{\rm m}(\rv)$ in matter, one finds a radiated energy
$dW/dL=(16\pi\alpha^2/9m^2) \, \gamma^2 \, \langle \Ev_{\rm m}^2 \rangle$,
which does not depend on the trajectory angle, apart from channeling effects.
The drawback of this classical prediction comes from applying (\ref{reglesom}) to a spectrum which overlaps the non-classical region ($\omega\sim\energ$), where quantum recoil corrections inhibits the radiation, not to speak of the forbidden region ($\omega>\energ$). 
Schematically, we can write 
\be
\Fv[\rv(t)] = \Fv_{\rm slow}(t) + \Fv_{\rm fast}(t) \,.  
\ee
where $\Fv_{\rm slow}$ and $\Fv_{\rm fast}$ generate the soft ($\omega\ll\energ$) and hard ($\omega\sim\energ$) radiations respectively. Equation (\ref{Larmor}) applies only to $\Fv_{\rm slow}$, giving the soft photon contribution $W_{\rm soft}$. When applied to $\Fv_{\rm fast}$, it overestimates $W_{\rm hard}$. For quasi-straight trajectories,
$\langle \Fv_{\rm slow}^2\rangle + \langle \Fv_{\rm fast}^2\rangle = \langle \Fv^2\rangle$ is the same at random or aligned
 crystal orientation, but not $\langle \Fv_{\rm slow}^2\rangle$ and $\langle \Fv_{\rm fast}^2\rangle$ separately. In the aligned case $\Fv_{\rm slow}$ comes essentially from the Lindhard potential, giving large $\langle \Fv_{\rm slow}^2\rangle$ and $W_{\rm soft}$, while $\Fv_{\rm fast}$ comes from the \emph{residual} atomic potentials. 
At random orientation, $\langle \Fv_{\rm slow}^2\rangle$ and $W_{\rm soft}$ are small; $\langle \Fv_{\rm fast}^2\rangle$ is larger than in the aligned case, but $W_{\rm hard}$ cannot increase much because of the recoil corrections. 

Nevertheless, Eq.(\ref{Larmor}) has something to tell in CBR~: when $\psi$  increases the CBR spectral peaks move together toward larger $\omega$, while the energies under the peaks remain constant \cite{GOUANERE}. Thus the total CBR energy does not depend on $\psi$. In fact it can be calculated by putting the Lindhard force in Eq.(\ref{Larmor}).

\section{Hard photon emission}

When the typical photon frequency discussed above is comparable to $\energ$, the classical theory over-estimates the radiated energy. This happens in a fast varying field, for instance in ordinary Bremsstrahlung and hard Compton scattering. It can also happen in uniform but very strong fields.  

\paragraph{Channeling radiation in QED-strong field.}

In an aligned crystal, a relativistic electron "sees" in its proper frame the field
$\Ev\T^* = \gamma \, \Ev\T$, where $ \Ev\T$ is the Lindhard field. At high $\gamma$, $\Ev\T^*$ can reach the critical QED value $ {\rm E}_{\rm crit}=m^2/e=1.32 \, 10^{18}$ volt/metre. 
The critical parameter is $\chi=\Ev\T^* / {\rm E}_{\rm crit}$.
For instance, $\chi\sim1$ is obtained at $\gamma\sim10^5$ along the $\langle110\rangle$ axis in germanium. 
In a critical electric field the Schwinger process of spontaneous pair creation should takes place, but in the channeling case it is prevented by the equally strong magnetic field $\Bv^* = \gamma \, \vv\times\Ev\T^*$. Nevertheless,  the following non-perturbative QED processes take place at $\chi\gtrsim1$ \cite{KIMBALL,BAIER} : 

\medskip \ni - "photon decay" $\gamma\to e^+ \, e^-$,  first proved in \cite{BELKACEM-split} (here $\chi\equiv(\omega/m) \, \Ev\T / {\rm E}_{\rm crit}$);

\medskip \ni - "photon splitting" $\gamma\to \gamma \, \gamma$ \cite{BAIER-split}.

\medskip \ni Besides, CR becomes "hard" ($\omega \sim \energ$). It leads to a large decrease or ``cooling" of the transverse energy $\energ\T$ at each emission, and, as a consequence, to a self-acceleration of the CR process \cite{BELKACEM, TIKHOMIROV, XA-cool,XA-code,BKS-cool}. The $\chi\sim1$ regime is also expected in beam-beam crossing at future $e^+e^-$ or $e^-e^-$ linear colliders, producing severe \emph{Beamsstrahlung} energy loss. 

\paragraph{The ``magic" Baier-Katkov (BK) formula.}

Equations (\ref{cov2}-\ref{noncov}) do not take into account the quantum recoil of the electron upon photon emission and the resulting spectrum may extend beyond the kinematic limit $\omega=\energ$. They are also blind to the electron spin. These defects are cured by rather simple modifications introduced by Baier and Katkov \cite{B-K}:

\medskip \ni - 1) for the recoil, replace $\phi = - k \ccdot X$ by $\Phi = - (\energ/\energ') \, k \ccdot X
= - (\energ/\energ') \, \omega\td$. 

\medskip \ni - 2) ${\cal A}$ becomes a spin-dependent amplitude $\langle s'|A_\epol |s\rangle$. In the helicity basis, 
\begin{subequations} \label{spinfacteur} 
\begin{align}
\langle+|A_-|+\rangle 
&=  \epol_-^* \cdot \int_{\cal T}  d\rv \, \exp(i\Phi) \,,
\quad
\langle-|A_+|-\rangle =  \epol_+^* \cdot \int_{\cal T} d\rv \,\exp(i\Phi) \,,
\\
\langle-|A_-|-\rangle 
&=  (\gamma/\gamma')  \, \langle+|A_-|+\rangle \,,
 \quad
\langle+|A_+|+\rangle =  (\gamma/\gamma')  \, \langle-|A_+|-\rangle \,,
\\
\langle-|A_+|+\rangle 
&= - \langle+|A_-|-\rangle =  2^{-1/2} \left(\frac{1}{\gamma'} - \frac{1}{\gamma} \right) \int_{\cal T} dt \,\exp(i\Phi) \,,
\\
\langle-|A_-|+\rangle 
&=  \langle+|A_+|-\rangle =  0 \,.
\end{align}
\end{subequations}
%
Note that  (\ref{spinfacteur}a) coincide with the classical formula except for $\phi\rightarrow\Phi$. 

The substitution $\phi\to\Phi$ can be understood using the sum-over-histories of Feynman:
the phase difference between the emission from two space-time points $X$ and $X+dX$ of the trajectory is
\be
d\Phi = (k+p'-p)\cdot dX \equiv  q\cdot dX \,, \quad \text{with } dX =  p \, dt/\energ \,,
\ee
instead of $d\phi=k\cdot dX$. The latter misses the propagation phase of the intial and final electrons. $q$ is the 4-momentum provided by the external field. Assuming a slow varying field, we can neglect $q^2$ and $\qv\t$ and derive
$k\cdot p\simeq q\cdot p'$ and $q\cdot p / q\cdot p' \simeq\energ/\energ'$, from where $d\Phi/d\phi=\energ/\energ'$. 

What seems "magic" is that the BK formula does not depend on the fate of the final electron, which may follow a very different trajectory. It only depends on the trajectory the electron would have without photon emission. 
 In fact, the BK formula can be obtained in the JWKB approach for an external potential which is invariant in the transverse coordinates, for instance in a counter-propagating plane wave. Then, by knowing the initial trajectory we know the final one (up to a lateral translation). 
Let us nevertheless test the BK formula in axial channeling, where the potential depends on transverse coordinates. Conservation of energy and longitudinal momentum gives
\be \label{raies1} 
({\omega/2}) \, \left[(\gamma\gamma')^{-1} + \theta^2\right] =  \energ\T - \energ'\T \,.
\ee
whereas, for a periodic trajectory of period $L$, the BK formula predicts 
\be  \label{raies2} 
({\gamma/\gamma'}) \, ({\omega/2}) \, \left(\gamma^{-2} + \theta^2+\langle\vv\T^2\rangle\right) = 2\nu\pi/{L} \,,
\ee
$\nu$ being the harmonic number. Let us assume that the transverse motion is quasi-classical and all channeled trajectories are periodic. Then $\nu=n-n'$, where $n\gg1$ is the principal quantum number given by the Bohr rule $L\energ\langle\vv\T^2\rangle=2 n \pi$. 
For a $\omega\ll\energ$ and $\nu\ll n$, we have 
\be
\energ\T - \energ'\T=2\nu\pi/L-\omega\langle\vv\T^2\rangle \,
\ee
(the last term comes from applying variational principle to the transverse Hamiltonian $V(\rv\T)+\pv\T^2/(2\energ)$ when $\energ$ changes). Then (\ref{raies1}) and  (\ref{raies2}) agree. However they give different results  when $\omega\sim\energ$ or $\nu\sim n$. Note that $\nu\ll n$ usually implies $\omega\ll\energ$, therefore we can conclude that the BK spectral lines are shifted only for the high harmonics. But it does not matter, since the latter build a quasi-continuum. 

The BK formula can be applied to linear and nonlinear Compton scattering, and to ordinary BR by factorizing the scattering amplitude and the radiation amplitude. This allows, in simulations \cite{XA-code}, to treat simultaneously the Channeling Radiation and the incoherent Bremsstrahlung.

\section{Impact parameter description}

\paragraph{Radiation as a tunneling effect.}

In vacuum, the reaction $e^-(p) \to \gamma(k) + e'^-(p')$ is kinematically forbidden if we consider it as a \emph{local} process between \emph{classical} particles. One cannot satisfy $p=p'+k$, $p^2=p'^2=m^2$, and $k^2=0$ together. 
In an external field, we can keep the particle description, assuming a \emph{nonlocal} process, where the $e'^-$ trajectory is at distance 
\be
\delta \rv = 
- \frac{\omega}{2\energ'} \,  \Rv \, (\gamma^{-2}+\theta^2) \,,
\ee
from the $e^-$ one,  
$\Rv$ being the $e^-$ position measured from the curvature centre.
This \emph{electron side slipping} \cite{XA-ss} is of the order of $|\delta \rv| \sim \omega/(m \,\omega_c) $ $\lesssim 400$ fermi, where $\omega_c=\gamma^3/R$ is the SR cutoff. In the the $\hbar\to0$ limit, side-slipping becomes continuous like the radiative energy loss. It gives the Schott term in radiation damping theory. 

To conserve angular momentum, the electron side-slipping is counterbalanced by the photon impact parameter (in ray optics)
\be
\bv = - (\energ'/\omega) \,  \delta \rv = \Rv \, (\gamma^{-2}+\theta^2)/2 \,.
\ee
Then, the $(e'^-+\gamma)$ center-of-energy prolongates the intial trajectories for some time (see Fig.1, right). 
For $\theta=0$ the photon trajectory is tangent to the \emph{light cylinder} of radius $R/v\simeq R(1+\gamma^{-2}/2)$. 
Outside this cylinder the Coulomb field of the electron would move faster than light.    
In reality, due to the wave nature of radiation, the impact parameter profile of Synchrotron Radiation contains fringes \cite{XA-impact}, like that of a caustic. This could be observed with a narrow enough electron beam. 

The ``photon side-slipping" of path $\bv$ can be considered as a tunneling effect. For a circular electron orbit, the photon has a angular momentum $\omega R/(v\cos\psi)$, where $\psi$ is defined below Eq.(\ref{Lcohcirc}). It must cross a centrifugal barrier which ends at the radius $R/(v\cos\psi)$. The above-cutoff behaviour %
\be
\sim \exp\left\{-(\omega/\omega_c) \, (1+\gamma^2\psi^2)^{3/2}
\right\}
\ee
of SR can be simply derived from the standard tunneling factor.
A similar tunneling effect takes place in the crossed process $\gamma\to e^+ e^-$ in an external field \cite{XA-blocking}. The impact relative impact parameter (see Fig.1, right) is 
\be
\bv(e^+) - \bv(e^-) = m^2_{\rm pair}  \, {\Fv\T} / ({2\omega \Fv\T^2}) \,,
\ee
where $\pm\Fv\T$ is the tranverse force acting on $e^\pm$.  

\section{Summary}

 We have presented, relying on significant equations, a few phenomenological aspects of radiation by relativistic electrons  in external field or in matter~: annular lobes with infrared-divergence,  constructive or destructive longitudinal coherence effects, 
release of the equivalent photons, semi-bare electrons, shadowing, quantum recoil and spin effects in ``artificial" strong fields, electron side-slipping, photon impact parameter and tunnelling mechanism. 

We did not broach many other important topics, for instance 
kinematical versus dynamical approaches of PXR, density effect, ionization energy loss, coherent emission from a whole electron bunch, stimulated emission, etc.  
They would deserve several other reviews.

%
%

\end{document}